\documentclass[aps,prb,twocolumn,amsmath,amssymb,gallery,footinbib,superscriptaddress,floatfix]{revtex4-2}
\usepackage{graphicx}
\usepackage{dcolumn}
\usepackage{bm}
\usepackage{hyperref}
\usepackage{stackrel,amssymb}
\usepackage{placeins}
\hypersetup{colorlinks=true, linkcolor=blue, citecolor=blue, urlcolor=magenta, pdftitle={On the search of displacive chiral phase transitions}, pdfauthor={F. Gomez-Ortiz}}
\usepackage{bbding}
\usepackage{xcolor} 
\usepackage[normalem]{ulem}

\begin{document}
\preprint{APS/123-QED}
\title{Mechanical coupling of polar topologies and  oxygen octahedra rotations in PbTiO$_3$/SrTiO$_3$ superlattices}
\author{Fernando G\'omez-Ortiz}
\email[Corresponding author:]{fgomez@uliege.be}
\affiliation{Theoretical Materials Physics, Q-MAT, Université de Liège, B-4000 Sart-Tilman, Belgium} 
\author{Louis Bastogne}
\affiliation{Theoretical Materials Physics, Q-MAT, Université de Liège, B-4000 Sart-Tilman, Belgium}
\author{Xu He}
\affiliation{Theoretical Materials Physics, Q-MAT, Université de Liège, B-4000 Sart-Tilman, Belgium}
\author{Philippe Ghosez}
\email[Corresponding author:]{Philippe.Ghosez@uliege.be}
\affiliation{Theoretical Materials Physics, Q-MAT, Université de Liège, B-4000 Sart-Tilman, Belgium} 

\date{\today}
\begin{abstract}
PbTiO$_3$/SrTiO$_3$ artificial superlattices recently emerged as a prototypical platform for the emergence and study of polar topologies. While previous studies mainly focused on the polar textures inherent to the ferroelectric PbTiO$_3$ layers, the oxygen octahedra rotations inherent to the paraelectric SrTiO$_3$ layers have attracted much little attention. Here, we highlight a biunivocal relationship between distinct polar topologies -- including $a_1/a_2$ domains, polar vortices, and skyrmions -- within  the PbTiO$_3$ layers and specific patterns of oxygen octahedra rotations in the SrTiO$_3$ layers. This relationship arises from a strain-mediated coupling between the two materials and is shown to be reciprocal. 
Through second-principles atomistic simulations, we demonstrate that each polar texture imposes a corresponding rotation pattern, while conversely, a frozen oxygen octahedra rotation dictates the emergence of the associated polar state. This confirms the strong coupling between oxygen octahedra rotations in SrTiO$_3$ and polarization in PbTiO$_3$, highlighting their cooperative role in stabilizing complex polar textures in related superlattices.
\end{abstract}
\maketitle
\section{Introduction}
\label{sec:introduction}
Over the past decade, complex polar textures in ferroelectric materials have attracted increasing attention from the solid-state physics community. Breakthroughs in epitaxial growing techniques~\cite{Dawber-05,Schlom-07,Mannhart-10}, which now allow atomic-layer precision, have enabled the fabrication of high-quality thin films and superlattices, where interfacial coupling and confinement give rise to emergent phenomena absent in bulk counterparts~\cite{Dawber-05,Bousquet-08}. These artificially structured systems have revealed a rich variety of nontrivial polar configurations~\cite{Junquera-23}, including a$_1$/a$_2$ domains~\cite{Pertsev-95,Damodaran-17,Celine-23}, flux-closure patterns~\cite{Jia-11,Tang-15}, vortices~\cite{Yadav-16}, electric skyrmion bubbles~\cite{Nahas-15,Pereira-19,Das-19}, merons~\cite{Nahas-20.2,Wang-20,Shao-23}, and hopfions~\cite{Lukyanchuk-20} among others~\cite{Stoica-19}, demonstrating how artificial layering can affect the polarization ground state.

Among the most extensively studied and celebrated platforms for exploring such emergent polar phenomena are superlattices composed of alternating layers of ferroelectric PbTiO$_3$ and dielectric SrTiO$_3$, owing to its structural simplicity, well-understood chemistry, and the rich variety of polar textures it supports. 
PbTiO$_3$/SrTiO$_3$ superlattices have been shown to host a wide range of topological polarization states depending on the interplay between electrostatic energies, elastic contributions, and gradient terms~\cite{Lines-01,Catalan-12,Junquera-23}. 
In most cases, SrTiO$_3$ is assimilated to a conventional dielectric and SrTiO$_3$ layers have been regarded as playing a secondary or passive role in these superlattices. Their primary function has been associated with the modification of the electrostatic boundary conditions experienced by the adjacent ferroelectric PbTiO$_3$ layers, forcing the system to break into domains~\cite{DeGuerville-05,Luk-09,Catalan-12,LUKYANCHUK2025}. In particular, the thickness of the SrTiO$_3$ dielectric layers has been used as a tuning parameter to modulate the strength of the depolarization field, thereby controlling whether the system behaves in an electrostatically coupled or decoupled regime~\cite{Bousquet-10,Stephenson-06,Zubko-11,Zubko-12}. For low thickness ratios of the dielectric material, the system operates in an electrostatically coupled regime, where the dielectric layer is expected to polarize with nearly the same spontaneous out-of-plane polarization as the adjacent ferroelectric layer. This results in a roughly uniform polarization across the entire structure. However, when the dielectric layer becomes sufficiently thick, the system transitions to an electrostatically decoupled regime, in which the polarization is largely confined within the ferroelectric layer.

Within this framework,  the oxygen octahedra rotations inherent to the antiferrodistorted bulk ground state of SrTiO$_3$ have recieved very little attention.
This passive picture has been further reinforced by the experimental difficulty in detecting octahedral rotations in SrTiO$_3$. Moreover, these rotations are known to disappear at elevated temperatures~\cite{Cowley-64,Shirane-69}, further limiting their perceived role in functional behavior.
Only in the low-thickness limit the coupling between the oxygen octahedra  rotations on the SrTiO$_3$ and the polar modes on PbTiO$_3$ has been considered~\cite{Bousquet-08,Aguado-11}. However, both the couplings and their coexistence are expected to vanish once the layer thickness exceeds three unit cells~\cite{Aguado-11}. 

In this work, we challenge the current assumptions and demonstrate that SrTiO$_3$ plays a far more active role in shaping the polar landscape of PbTiO$_3$/SrTiO$_3$ superlattices than previously recognized. Specifically, we show that there exists a mechanical coupling between both materials. This mechanical coupling leads to a biunivocal relationship between specific oxygen octaedra rotation patterns in SrTiO$_3$ and distinct polar textures in PbTiO$_3$, identifying SrTiO$_3$ as an active participant in the stabilization and selection of polar topologies.
Such coupling schemes could be harnessed to modulate emergent functionalities such as magnetoelectric responses in multiferroic heterostructures where polar states correlate with rotation patterns that, in turn, can couple to the spin order~\cite{Benedek-11}.
\begin{figure}[thbp]
    \centering
      \includegraphics[width=6cm]{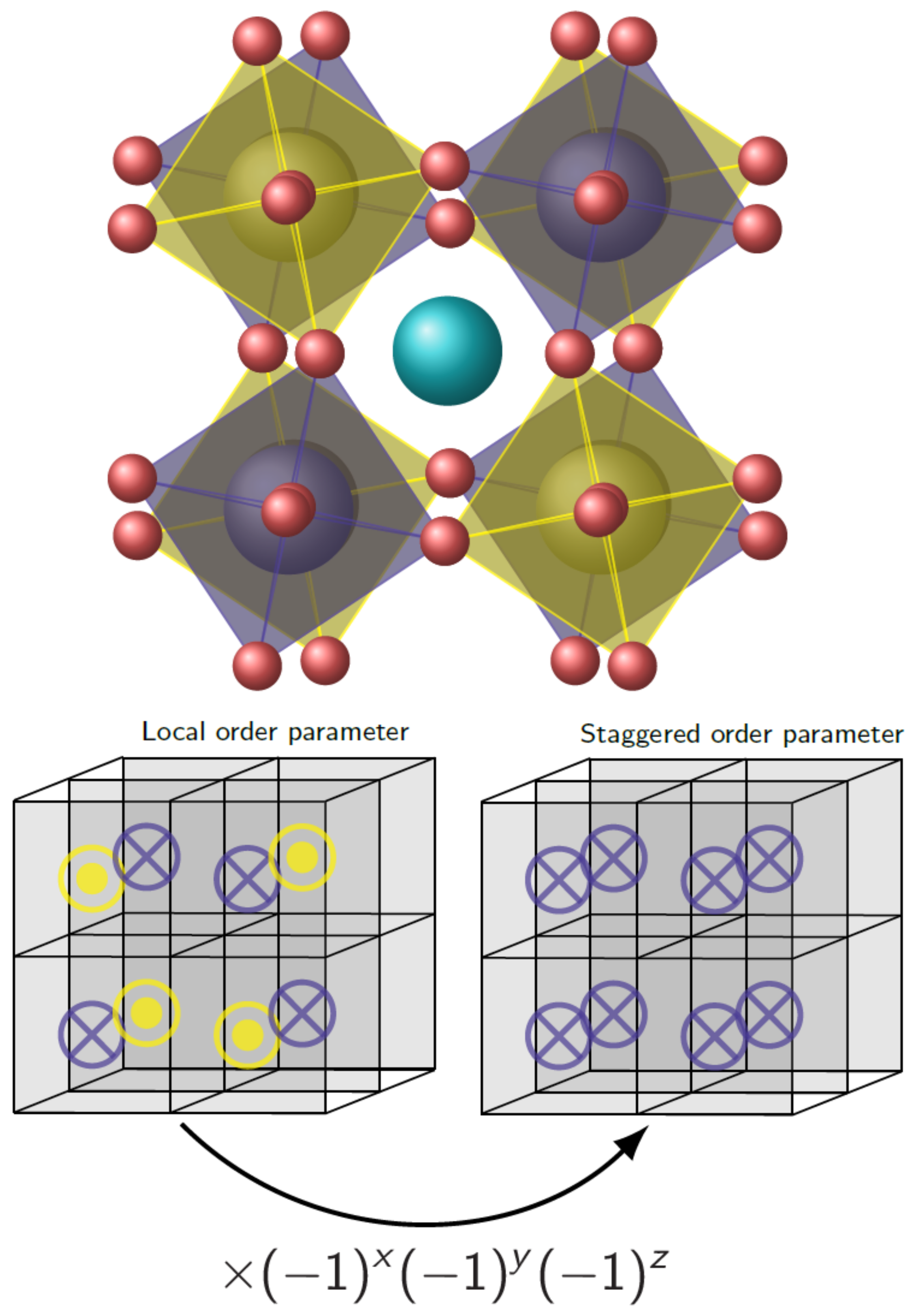}
      \caption{Schematic illustration of oxygen octahedra rotations following an out-of-phase antiferrodistortive mode along  an axis perpendicular to the plane of the paper in a perovskite structure. The pristine and the staggered local rotations of the TiO$_6$ octahedra are shown.}
      \label{fig:sketch_rot} 
\end{figure}
\section{Methods}
\label{sec:Methods}
Second-principles (SP) atomistic models were constructed using the \textsc{Multibinit}~\cite{gonze2020abinit,abinit2025} software by fitting data from density functional theory (DFT) calculations produced with the \textsc{Abinit}~\cite{gonze2020abinit,abinit2025} software package. PBESol exchange-correlation functional and a planewave-pseudopotential approach with optimized norm-conserving pseudopotentials from the PseudoDojo server~\cite{hamann2013optimized,van2018pseudodojo} were employed.
After constructing the bulk compound models~\cite{Bastogne2025}, the superlattice models were derived using a standard approach, as described in Ref.~\cite{Zubko-16}. The resulting superlattice models are identical to those employed in Ref.~\cite{Zatterin-24}, where further technical details are provided.

Structural relaxations were carried out using a combination of Hybrid Molecular Dynamics-Monte Carlo approach~\cite{duane1987hybrid,betancourt2017conceptual} at very low temperatures (T$=1$ K) together with Broyden-Fletcher-Goldfarb-Shanno (BFGS) algorithm as implemented in {\sc{abinit}}~\cite{gonze2020abinit,abinit2025}.

In order to compute the polarization profile, we followed the regular approach where local polarizations are computed within a linear approximation as the product of the Born effective charge tensor (in the reference bulk cubic phase of each compound) times the atomic displacements from the reference structure positions. An average over a volume of one unit cell centered at the Pb/Sr sites was used. Therefore, the displacements taken into account involve the given Pb/Sr site, its eight nearest Ti and its 12 nearest oxygen neighbors. The resulting value was divided by the volume of the unit cell~\cite{meyer2002ab}.

Oxygen octahedra rotations are visualized as arrows representing the rotation of the TiO$_6$ octahedra, with each arrow centered on Ti and indicating the local rotation axis and the magnitude of the tilt as schematized in Fig.~\ref{fig:sketch_rot}. 
Since oxygen octahedra rotations are antiferrodistortive motions involving anti-phase rotations in neighboring unit cells, we define and plot a {\it  staggered} rotation field, obtained by multiplying the rotation in unit cell ($x, y, z$) by a phase correction factor of $(-1)^x(-1)^y(-1)^z$ (see Fig.~\ref{fig:sketch_rot}). This transformation allows a uniform AFD rotation pattern to appear as a monodomain state for clearer visualization. Note that this phase convention is origin-dependent, and as a result, the plotted rotation field is only defined up to a global sign.

\begin{figure*}[tbh]
     \centering
      \includegraphics[width=\textwidth]{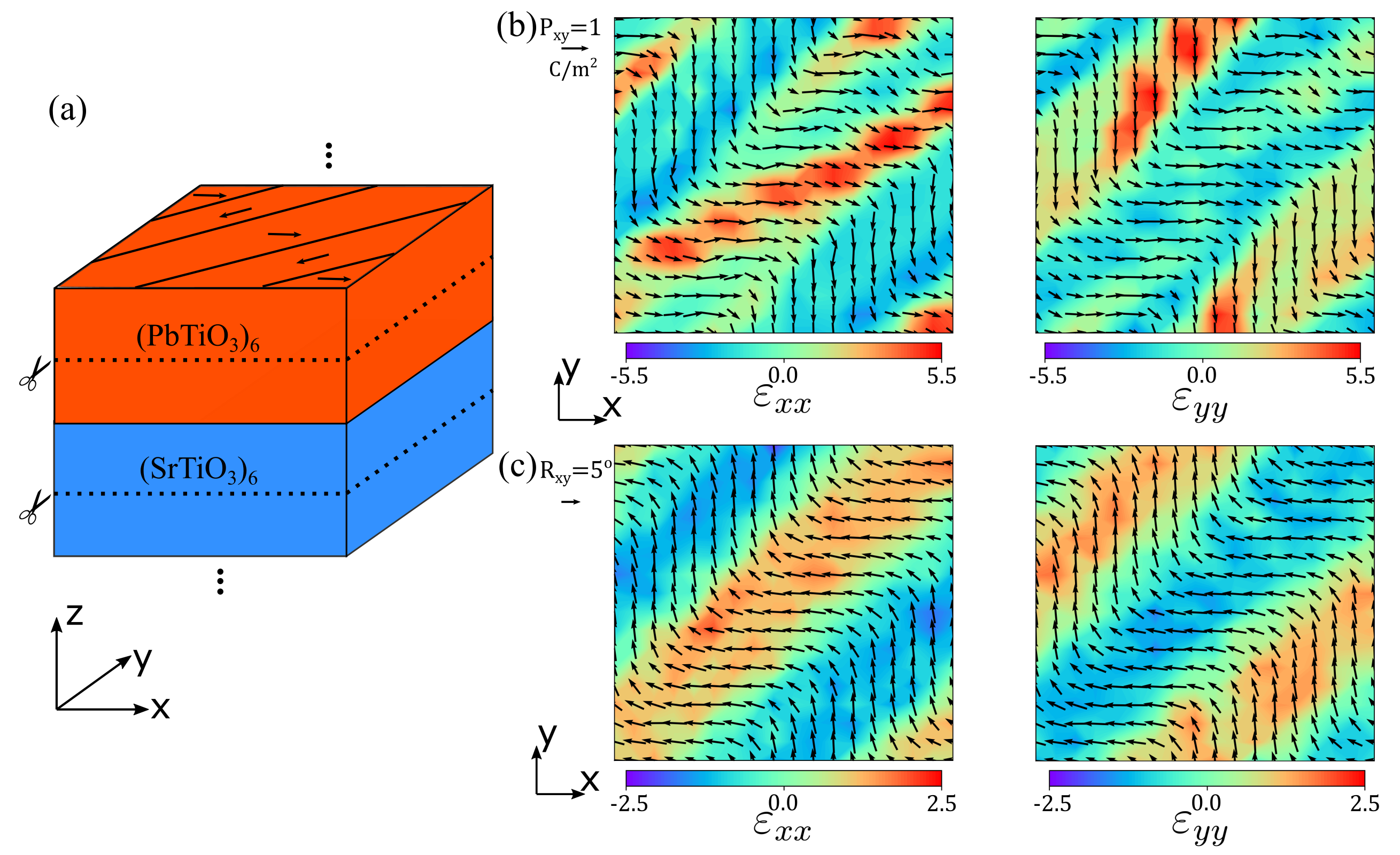}
      \caption{(a) Schematic representation of the (PbTiO$_3$)$_{6}$/(SrTiO$_3$)$_{6}$ used for the relaxation of the $a_1$/$a_2$ phase. (b) Planar $xy$-view of the polarization profile on a central PbTiO$_3$ layer as depicted by the dashed line and scissors. (c) Planar $xy$-view of the antiferrodistortive profile on a central SrTiO$_3$ layer as depicted by the dashed line and scissors. In both panels, arrows represent the in-plane components of the corresponding vector fields. The scale for arrow length is indicated on each panel in units of C/m$^2$. The color map denotes the local in-plane strain components along the $x$ or $y$ directions as indicated by the labels on the colorbar computed in \% taking $3.917$\AA~as a reference.}
      \label{fig:figa1a2} 
\end{figure*}
\section{Results}
In the following, we explore the most representative polarization textures that emerge in PbTiO$_3$/SrTiO$_3$ superlattices, focusing on the a$_1$/a$_2$ domain structure~\cite{Damodaran-17,Celine-23}, polar vortices~\cite{Tang-15,Yadav-16,Celine-23}, and polar skyrmions~\cite{Zhang-17,Das-19}. These configurations have been widely reported in the literature~\cite{Junquera-23}. 
%

\emph{The a$_1$/a$_2$ phase} has been experimentally observed in symmetric superlattices at low periodicities when they are grown on top of DyScO$_3$ substrates ($a=3.947$\AA~$b=3.951$\AA), which imposes only tiny strains -0.25\%--0.16\% on cubic bulk PbTiO$_3$~\cite{Damodaran-17,Celine-23}. In our simulations, it can be stabilized and results in a metastable phase when imposing 0\% epitaxial strain with respect to the cubic PbTiO$_3$ reference. Specifically, in Fig.~\ref{fig:figa1a2} we show the schematic representation of the  $18\times18\times\left[\right.$(PbTiO$_3$)$_{6}$/(SrTiO$_3$)$_{6}$$\left .\right]$ used in the calculations under the epitaxial strain condition of $3.917$ \AA~ that corresponds to the bulk value of the cubic phase predicted by the model for the PbTiO$_3$ crystal~\cite{Bastogne2025}.
\begin{figure*}[tbh]
     \centering
      \includegraphics[width=\textwidth]{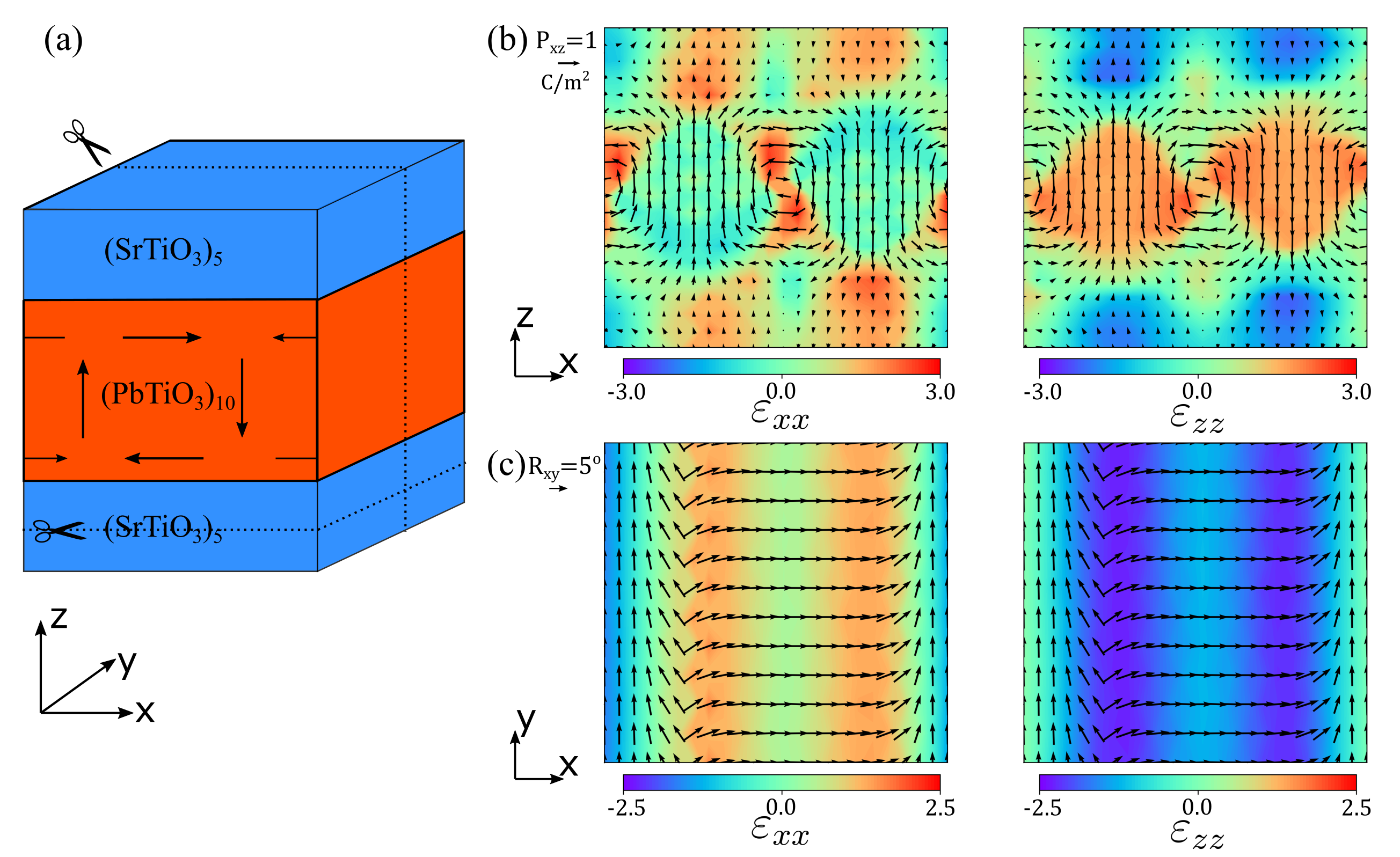}
      \caption{(a) Schematic representation of the (PbTiO$_3$)$_{10}$/(SrTiO$_3$)$_{10}$ used for the relaxation of the polar vortex phase. (b) Front $xz$-view of the polarization profile on the supperlattice as schematized by the dashed line and scissors. (c) Planar $xy$-view of the antiferrodistortive profile on a central SrTiO$_3$ layer as schematized by the dashed line and scissors. In both panels, arrows represent the in-plane components of the corresponding vector fields in C/m$^2$ and degrees respectively. The scale for arrow length is indicated on each panel. The color map denotes the local in-plane strain components along the $x$ or $z$ directions as indicated by the labels on the colorbar computed in \% taking $3.917$\AA~as a reference. For visualization purposes, local strain values in the PbTiO$_3$ layers are divided by a factor of 2 to allow both PbTiO$_3$ and SrTiO$_3$ to be shown on a common color scale without oversaturating the plot and to preserve visibility in the SrTiO$_3$ region.}
      \label{fig:figvortex} 
\end{figure*}
As shown in Fig.~\ref{fig:figa1a2}(b), after relaxation, the polarization pattern adopts a domain structure to accommodate the imposed macroscopic strain conditions with regions showing alternating $x$- and $y$-directed polar domains showing polarization magnitudes of about $0.9$ C/m$^2$ separated by domain walls along $\lbrace110\rbrace_{\rm{pc}}$ direction. These regions are associated with ferroelastic local strain domains. When the polarization is directed along the $x$-axis the system exhibits local expansive strain along $x$ and local compressive strain along $y$, while regions with $y$-oriented polarization show the opposite behavior i.e. expansive along $y$ and compressive along $x$ as shown by the color maps of Fig.~\ref{fig:figa1a2}(b). Such local strain distributions naturally arise as the system accommodates to the bulk values.

Remarkably, the strain state in the PbTiO$_3$ layers is mechanically coupled to the adjacent SrTiO$_3$ layers, which adopt a compatible -- although smaller --deformation pattern as shown in Fig.~\ref{fig:figa1a2}(c). As a result, the SrTiO$_3$ layers mirror the a$_1$/a$_2$ domain configuration in terms of the antiferrodistortive octahedral rotation pattern that follow the same domain periodicity.

Therefore, the $a_1$/$a_2$ phase exhibits a strong correlation between the AFD octahedral rotations on the SrTiO$_3$ layers and the polarization pattern on the PbTiO$_3$ layers, arising from the emergence of ferroelastic domains and mechanical coupling between the two compounds.

\emph{The polar vortex phase} and flux closure domains have been experimentally observed in a large range of layer thicknesses under slightly tensile epitaxial strain~\cite{Tang-15,Yadav-16,Celine-23}. In our simulations, it can be stabilized and results a metastable phase in a large window of epitaxial strain and layer thicknesses as demonstrated in Ref.~\cite{Zatterin-24}. In this work, we focus ourselves on the (PbTiO$_3$)$_{10}$/(SrTiO$_3$)$_{10}$ case grown under the epitaxial strain condition of $3.917$ \AA. Under these conditions, the optimal domain periodicity predicted by our model is 24 unit cells~\cite{Zatterin-24}. We simulate a $24\times6\times[\mathrm{(PbTiO_3)_{10}/(SrTiO_3)_{10}}]$ supercell, where 6 unit cells along the $y$-direction are sufficient since the vortex tubes are periodic and extend along that direction, and allow the TiO$_6$ octahedral rotations to relax without artificial constraints. This choice ensures that octahedral tiltings are accurately represented, while also keeping the computational cost tractable.

As shown in Fig.~\ref{fig:figvortex}(a),  vortex tubes extend along the axial $y$-direction and exhibit a periodic alternation of clockwise and counterclockwise rotation along the $x$-direction. Regions of ``up'' ($+z$) and ``down'' ($-z$) polarization are connected
by flux closure domains where the polarization lays along the $x$-direction to minimize the stray fields at the interface. As a consequence, local strain regions with dominant expansive strain along the $z$ direction are encountered at the domain regions whereas local regions showing a positive $\varepsilon_{xx}$ are encountered at the flux-closure regions near the interface where the vortex develop as shown in the color maps of Fig.~\ref{fig:figvortex}(b). 
\begin{figure*}[tbh]
     \centering
      \includegraphics[width=\textwidth]{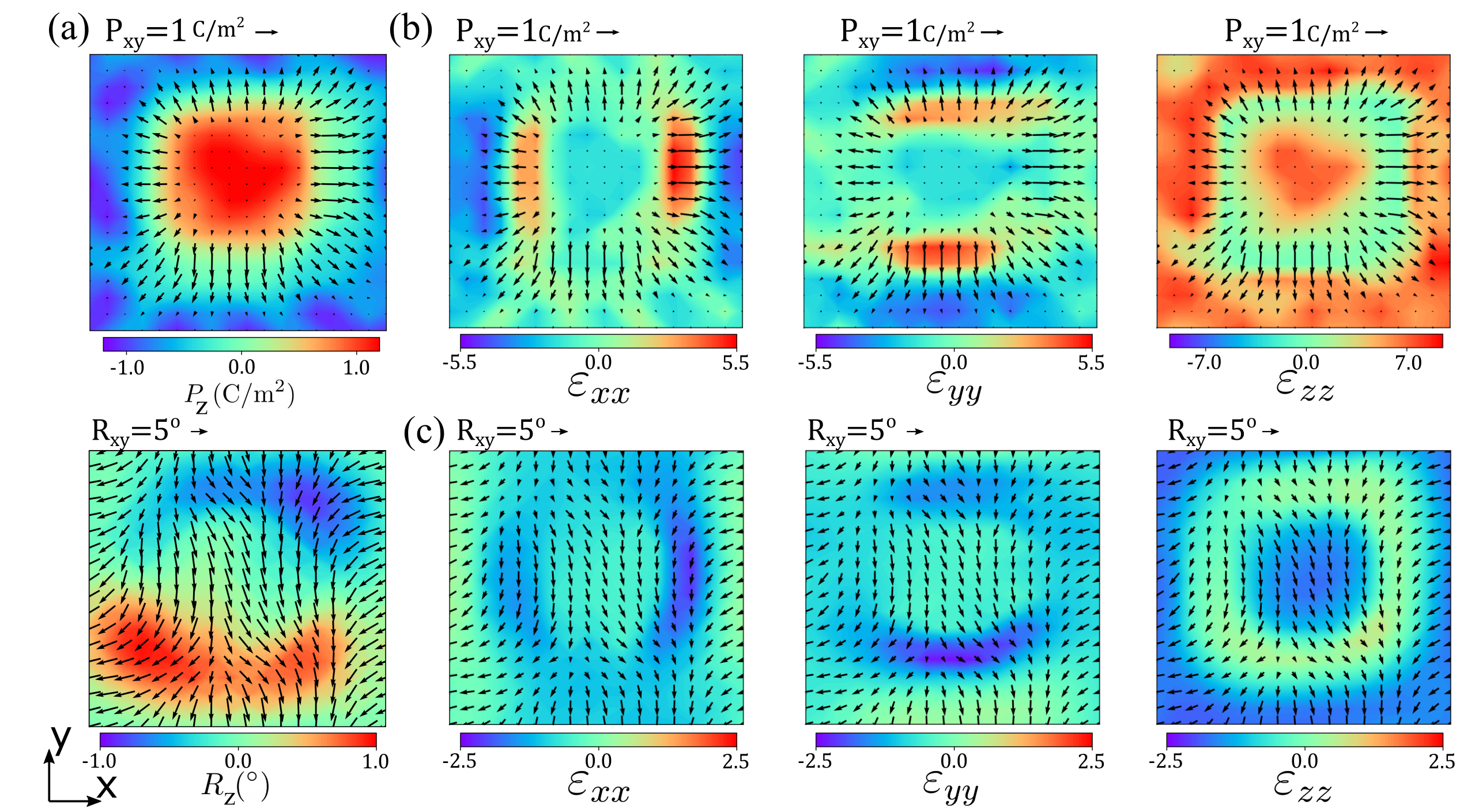}
      \caption{Polar bubble phase on the (PbTiO$_3$)$_{6}$/(SrTiO$_3$)$_{6}$ superlattice. (a) Planar $xy$-view with the polarization map on a PbTiO$_3$ layer near the interfase and antiferrodistortive map on the SrTiO$_3$ on a central layer. Color maps indicate the poalrization and antiferrodistortive rotations along the $z$-direction. (b) Planar $xy$-view of the in-plane polarization profile on the supperlattice. (c) Planar $xy$-view of the in-plane antiferrodistortive profile on a central SrTiO$_3$ layer. In both panels, arrows represent the in-plane components of the corresponding vector fields. The scale for arrow length is indicated on each panel. The color map denotes the local in-plane strain components along the $x$, $y$ and $z$ directions as indicated by the labels on the colorbar computed in \% taking $3.885$\AA~as a reference.    }
      \label{fig:figskr} 
\end{figure*}
Then, the local strains in the SrTiO$_3$ layers adopt the shape shown in Fig.~\ref{fig:figvortex}(c).
This pattern accommodates the strain components in the PbTiO$_3$ layers as well as the periodic boundary conditions. 
In particular, to satisfy periodicity, the average of the local strains along $i$-direction must equal the externally imposed macroscopic strain condition along that particular direction. 
As a result, dominant local expansive strains are directed along the $x$-directions and dominant local compressive strain components are encountered along $z$. 
Therefore, the antiferrodistortive rotations show a different pattern compared to the previous case that is invariant along the direction of the vortex tubes (as it is the case of the polarization and local strains) and show a predominant $x$-component compatible with the local strains as shown in Fig.~\ref{fig:figvortex}(c).

\emph{The polar bubble phase} or polar skyrmions have been experimentally stabilized under small compressive strains, typically grown on top of SrTiO$_3$ substrate~\cite{Zhang-17,Das-19,Junquera-23}. In our calculations we used a $18\times18\times\left[\right.$(PbTiO$_3$)$_{6}$/(SrTiO$_3$)$_{6}$$\left .\right]$ under the epitaxial condition of $3.887$ \AA~ which almost corresponds to the bulk cubic lattice parameter predicted for SrTiO$_3$ by the model ($3.889$ \AA). Under these conditions the polar bubble phase is metastable showing a characteristic columnar nanodomain of polarization with in-plane center convergent and divergent patterns at the top and bottom interfaces to counterbalance the out-of plane polarization divergence as shown in Fig.~\ref{fig:figskr}(a). From a mechanical point of view, the strains developed on the PbTiO$_3$ show predominant local tensile strains along the $z$-direction in both the nanocolumn and the matrix, with a localized region exhibiting zero to slightly negative local strain, forming a ``doughnut-like'' structure where local tensile strains along the $x$- and $y$-directions develops. 
Similar to the previous case, these mechanical conditions influence the strain state in the SrTiO$_3$ that shows a complementary doughnut-like structure for the $z$-strain to counterbalance those appearing on the PbTiO$_3$ and minimize the stress along the $z$-direction [see Fig.~\ref{fig:figskr}(c)]. As a result, the rotation pattern exhibits a predominantly in-plane structure, with an out-of-plane component that is five times smaller (though not negligible) in the doughnut-like region as shown in Fig.~\ref{fig:figskr}(a).

\section{Discussion}
So far, we have primarily stabilized the polarization pattern in the PbTiO$_3$ layers and analyzed its effect on the TiO$_6$ octahedral rotations of SrTiO$_3$ layers, demonstrating that the different polarization patterns ($a_1$/$a_2$, polar vortices and polar bubbles) do have an impact on the antiferrodistortive rotations displayed on the SrTiO$_3$. However, we now disucss that the opposite is equally true: the SrTiO$_3$ layers are not just passive recipients of PbTiO$_3$ polarization patterns, but can also actively contribute to shape the resulting polar configuration.
To demonstrate such a claim we conducted constrained relaxations where the positions of the oxygens on the SrTiO$_3$ are fixed throughout the relaxation to keep the previously obtained octahedral rotation patterns constant. In contrast, the atoms in the PbTiO$_3$ layers, as well as the Sr and Ti atoms in the SrTiO$_3$ layers, were initialized at their cubic reference positions, ensuring no initial polar distortion, and allowed to relax freely. 
After a few relaxation steps, the characteristic polarization patterns of the PbTiO$_3$ are recovered demonstrating that the correlation between the ferroelectric modes at the PbTiO$_3$ layers and the antiferrodistortive modes at the SrTiO$_3$ layers is reciprocal. 
This coupling persists even though the rotations in the SrTiO$_3$ do not penetrate the PbTiO$_3$ layers, which exhibit no antiferrodistortive mode suggesting that the driving force behind this behavior is the local strains and mechanical coupling among the layers. An explicit example of this procedure, including how the rotations are fixed during the simulation and how the polarization develops in PbTiO$_3$, is presented in the Appendix for the $a_1/a_2$ case.

\begin{figure}[tbh]
     \centering
      \includegraphics[width=8cm]{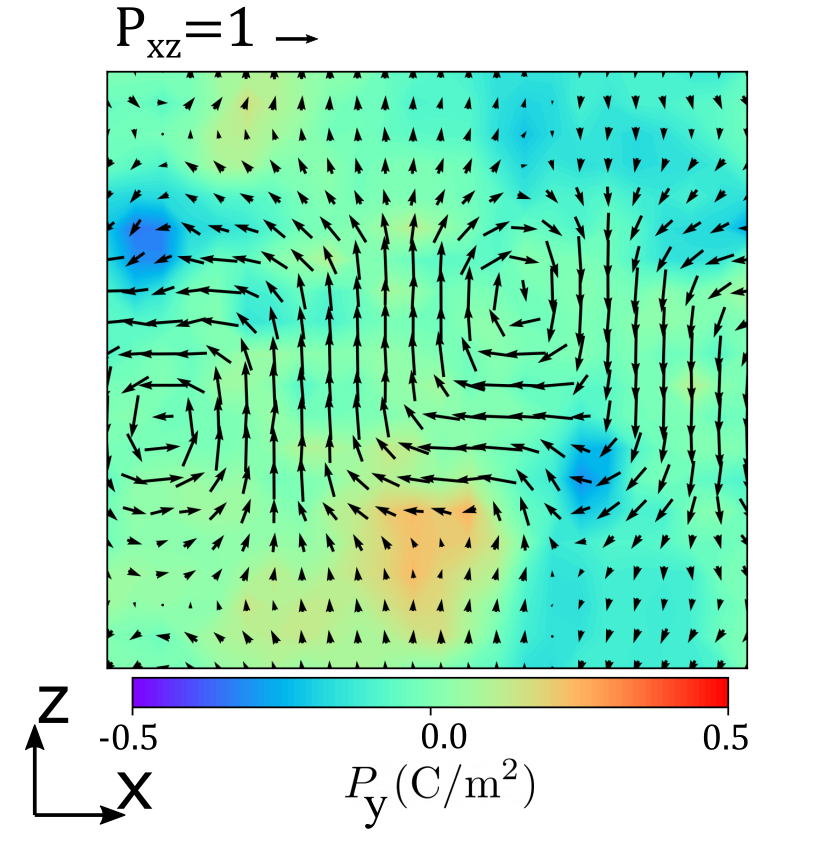}
      \caption{Polar vortex phase obtained after the constrained relaxation fixing the TiO$_6$ octahedral rotations on the SrTiO$_3$ to the values encountered on the first part of the manuscript. Clockwise counter-clockwise vortices are displaced towards the top/bottom interfaces and a net in-plane polarization along the $x$-direction is developped on the PbTiO$_3$}
      \label{fig:figbuck} 
\end{figure}

Remarkably, the only notable difference observed between the polarization patterns obtained after this constrained relaxation and the original a$1$/a$2$, polar vortex and polar bubble phases is that, in the polar vortex phase, the system shows a buckling of the vortices (Fig.~\ref{fig:figbuck}), indicating a subtle influence of the tensile strain along the $x$-direction in the SrTiO$_3$. This strain penetrates the PbTiO$_3$ enhancing a $P_x$ polarization component, which has been associated to the emergence of a buckling state~\cite{Behera-22,Behera-23}.
\section{Conclusions}
In this work we have demonstrated the coupling between the oxygen octahedra rotations in the SrTiO$_3$ layers and the polar distortions in the PbTiO$_3$ layers. By studying the most representative polar phases encountered in PbTiO$_3$/SrTiO$_3$ supperlattices (such as the a$_1$/a$_2$ domains, polar vortices, and polar bubbles), we have shown that each of these configurations is intimately linked to a unique and distinct antiferrodistortive rotation pattern in the SrTiO$_3$ layers. 
Even though, in agreement with previous works~\cite{Bousquet-08,Aguado-11}, there is no coexistence of oxygen octahedra rotations and polar distortions in PbTiO$_3$ layers for our investigated thicknesses, we have shown that the coupling between the two types of distortions in individual layers is driven by mechanical interactions and crosstalking between local strains.

Indeed, we have proven that fixing the TiO$_6$ octahedral rotations in the SrTiO$_3$ is enough to drive the system towards the desired polar phase. These findings challenge the prevailing view of the dielectric medium as a mere passive component whose primary role is to modulate depolarizing fields. Instead, our results underscore its active involvement in shaping the polarization texture, opening the door to novel mechanisms for the stabilization of topological states. This perspective becomes particularly compelling in multiferroic systems, where octahedral tilts are entangled with other order parameters such as magnetism~\cite{Benedek-11}, potentially enabling cross-coupled control of topological structures.

\acknowledgments
F.G.O. acknowledges financial support from MSCA-PF 101148906 funded by the European Union and the Fonds de la Recherche Scientifique (FNRS) through the grant FNRS-CR 1.B.227.25F and the Consortium des Équipements de Calcul Intensif (CÉCI), funded by the F.R.S.-FNRS under Grant No. 2.5020.11 and the Tier-1 Lucia supercomputer of the Walloon Region, infrastructure funded by the Walloon Region under the grant agreement No. 1910247. F.G.-O. and Ph. G. also acknowledge support by the European Union’s Horizon 2020 research and innovation program under Grant Agreement No. 964931 (TSAR).  Ph. G. and X.H. also acknowledge support from the Fonds de la Recherche Scientifique (FNRS) through the PDR projects PROMOSPAN (Grant No. T.0107.20) and TOPOTEX (Grant No. T.0128.25).
%
\clearpage
\onecolumngrid
\appendix
\renewcommand{\thefigure}{A\arabic{figure}}
\setcounter{figure}{0} 
\section{Constrained relaxation}
\begin{figure*}[bthp]
     \centering
      \includegraphics[width=\textwidth]{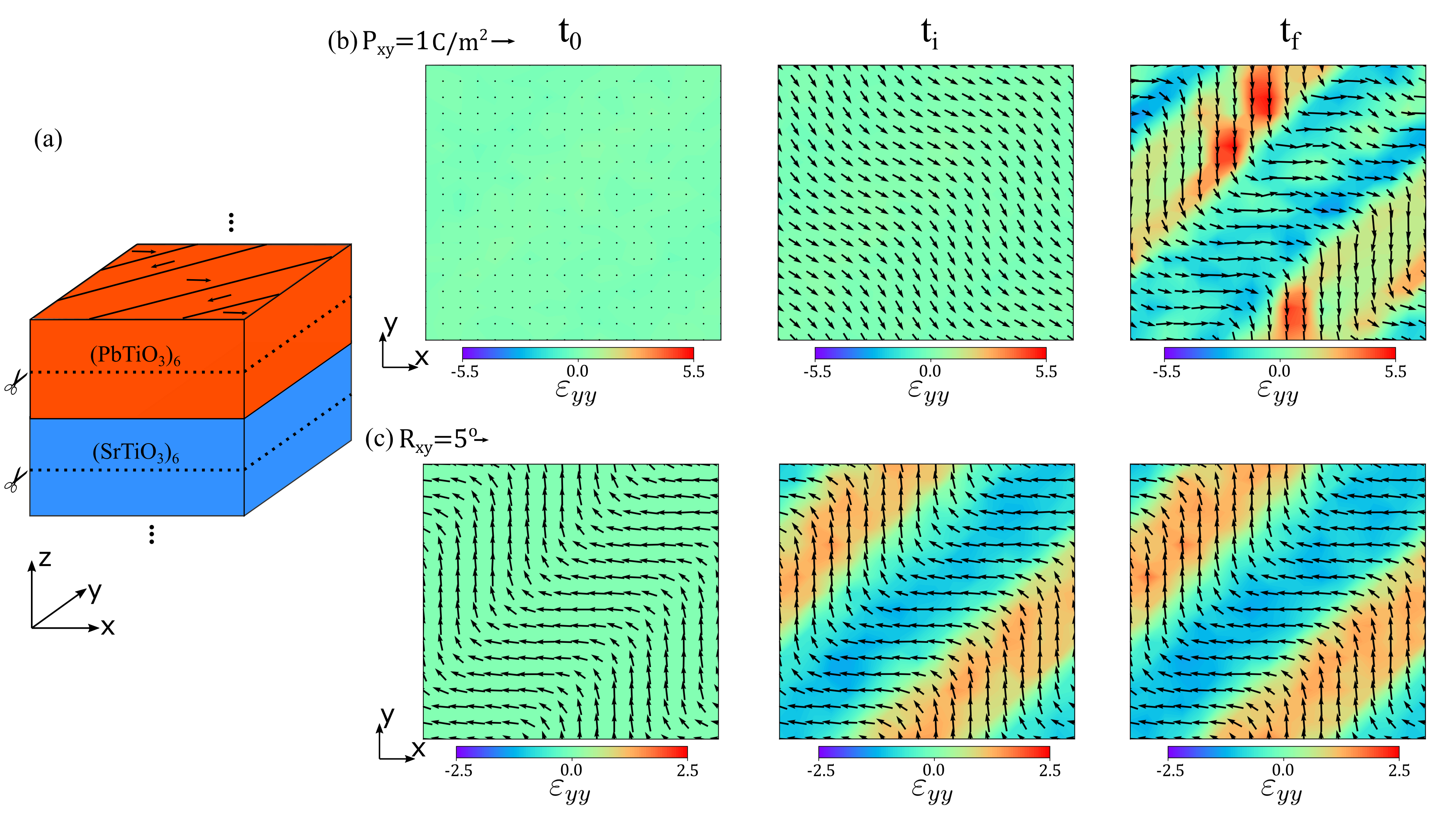}
      \caption{(a) Schematic representation of the (PbTiO$_3$)$_{6}$/(SrTiO$_3$)$_{6}$ used for the constrained relaxation of the $a_1$/$a_2$ phase explained in the main text. (b) Planar $xy$-view of the polarization profile on a central PbTiO$_3$ layer as depicted by the dashed line and scissors at different times of the constrained relaxation. (c) Planar $xy$-view of the antiferrodistortive profile on a central SrTiO$_3$ layer as depicted by the dashed line and scissors at different steps of the constrained relaxation. In both panels, arrows represent the in-plane components of the corresponding vector fields. The scale for arrow length is indicated on the bottom of each panel in units of C/m$^2$ and degrees respectively. The color map denotes the local in-plane strain components along the $y$ direction computed in \% taking $3.917$\AA~as a reference.}
      \label{fig:supp} 
\end{figure*}

In this section, we exemplify the constrained relaxation procedure described in the main text using the $a_1/a_2$ configuration as a representative case. 
During the relaxation the positions of the oxygens on the SrTiO$_3$ are fixed to keep the octahedral rotation patterns obtained in the main text constant. In contrast, the atoms in the PbTiO$_3$ layers, as well as the Sr and Ti atoms in the SrTiO$_3$ layers, were initialized at their cubic reference positions, ensuring no initial polar distortion, and allowed to relax freely.
Figure~\ref{fig:supp} shows that the octahedral rotations in SrTiO$_3$ remain unchanged throughout the relaxation steps. Initially, the local strain is zero, as the Sr atoms begin in centrosymmetric positions. However, they quickly adapt to the imposed rotation pattern, developing the relaxed local strain configuration. This strain propagates into the PbTiO$_3$ layers, leading to the recovery of the characteristic polarization pattern, as shown in Fig.~\ref{fig:figa1a2}.
\end{document}